\documentclass[conference]{IEEEtran}
\usepackage{graphicx}
\usepackage{stfloats}
\usepackage{cite}
\usepackage{graphicx}
\usepackage{verbatim}
\usepackage[cmex10]{amsmath}
\usepackage{multirow}

\ifCLASSINFOpdf
\else
\fi
\def\BibTeX{{\rm B\kern-.05em{\sc i\kern-.025em b}\kern-.08em
		T\kern-.1667em\lower.7ex\hbox{E}\kern-.125emX}}

\begin{document}

\title{Domain Priori Knowledge based 
Integrated Solution Design for Internet of Services}

\author{\IEEEauthorblockN{Hanchuan Xu, Xiao Wang, Yuxin Wang, Nan Li, Zhiying Tu, Zhongjie Wang, Xiaofei Xu}
\IEEEauthorblockA{School of Computer Science and Technology, Harbin Institute of Technology\\
Harbin, China\\
Email: \{xhc, wxlxq\}@hit.edu.cn, 19B903081@stu.hit.edu.cn, \{linan, tzy\_hit, rainy, xiaofei\}@hit.edu.cn}
}

\maketitle

\begin{abstract}
    Various types of services, such as web APIs, IoT services, O2O services, and many others, have flooded on the Internet. Interconnections among these services have resulted in a new phenomenon called ``Internet of Services'' (IoS). By IoS, people don't need to request multiple services by themselves to fulfill their daily requirements, but it is an IoS platform that is responsible for constructing integrated solutions for them. Since user requirements (URs) are usually coarse-grained and transboundary, IoS platforms have to integrate services from multiple domains to fulfill the requirements. Considering there are too many available services in IoS, a big challenge is how to look for a tradeoff between the construction efficiency and the precision of final solutions. For this challenge, we introduce a framework and a platform for transboundary user requirement oriented solution design in IoS. The main idea is to make use of domain priori knowledge derived from the commonness and similarities among massive historical URs and among historical integrated service solutions(ISSs). Priori knowledge is classified into three types: requirement patterns (RPs), service patterns (SPs), and probabilistic matching matrix (PMM) between RPs and SPs. A UR is modeled in the form of an intention tree (I-Tree) along with a set of constraints on intention nodes, and then optimal RPs are selected to cover the I-Tree as much as possible. By taking advantage of the PMM, a set of SPs are filtered out and composed together to form the final ISS. Finally, the design of a platform supporting the above process is introduced.  
\end{abstract}


% \fi
\IEEEpeerreviewmaketitle

%%2020-02-14 文字里用的“we”太多了，建议改成客观陈述的语态

\section{Introduction}

In the big data era, servitization becomes one of the most important development trends in the IT world. More and more software services are developed and deployed on the Internet, along with a huge number of virtualized services that connect real-world physical service resources. Services from multiple domains, multiple networks, and multiple worlds are converged as a huge complicated service network or ecosystem, which can be called as ``Internet of Services (IoS)''\cite{IoS} or ``Big Service''\cite{bigservice}. IoS presents a paradigm in which everything is available as a service on the Internet. In IoS, the extremely abundant massive services are diverse, distributed, and heterogeneous. By gathering, clustering, and composing these services, service solutions are produced to meet customer requirements. IoS is customer-focused so that when it receives a customer’s requirement, it creates integrated service solutions(ISSs) on demand, combining service resources to complete the service tasks. How to reuse the abundant service resources to rapidly develop new applications or ISSs to meet massive individualized customer requirements is the key issue in the IoS ecosystem.

%%2020-02-14 by WZJ: 下面这段，建议把所陈述的技术挑战问题用列表的形式给出来，显得清晰

%\begin{itemize}
%    \item 
%\end{itemize}
%% 另一个问题：感觉目前所列的已有工作，都是从0开始做建模和构造的，比较传统，跟咱们IoS背景下的服务方案构造问题有些远。可否增加一些关于服务大规模定制、服务网络定制、服务的大粒度复用、甚至服务模式方面的已有工作的论述？

Main technique challenges in IoS include:
\begin{itemize}
   \item There are too many available services in IoS. A big challenge is how to look for a tradeoff between the construction efficiency and the precision of final solutions.

    \item Facing massive individualized customer requirements, how to assist users to present their requirements accurately in an understandable and convenient manner, and also for efficient construction of service solution is the key challenge in requirement elicitation and modeling.

    \item IoS is a far complex ecosystem that consists of various organizations and domains, multi-platforms, massive services, and users. How to design and implement a platform to support the efficient service solution construction and operation in IoS is a challenge.

\end{itemize}

Researchers on service engineering and service computing have proposed some service developing paradigms, such as  Service Centric Systems Engineering (SeCSE)\cite{di2009secse}, Service Model Driven Architecture (SMDA)\cite{xu2007smda}, Service-Oriented Modeling and Architecture (SOMA)\cite{arsanjani2008soma} as well as thousands of approaches of service selection and composition\cite{sheng2014web} to address the issues of composite service design problem. There are also some researches for service mass customization\cite{brax2017service}, service network customization\cite{wang2013mass}, and large-grained reuse of services\cite{Rong2016Reusing}. However, there are still no sophisticated approaches to construct services accurately and efficiently, and developing service solutions to meet massive individualized customer requirements is still a very cumbersome and time-consuming activity, especially when these services are complex.

\begin{figure}[ht]
	\centering
	\includegraphics[width=3.5in]{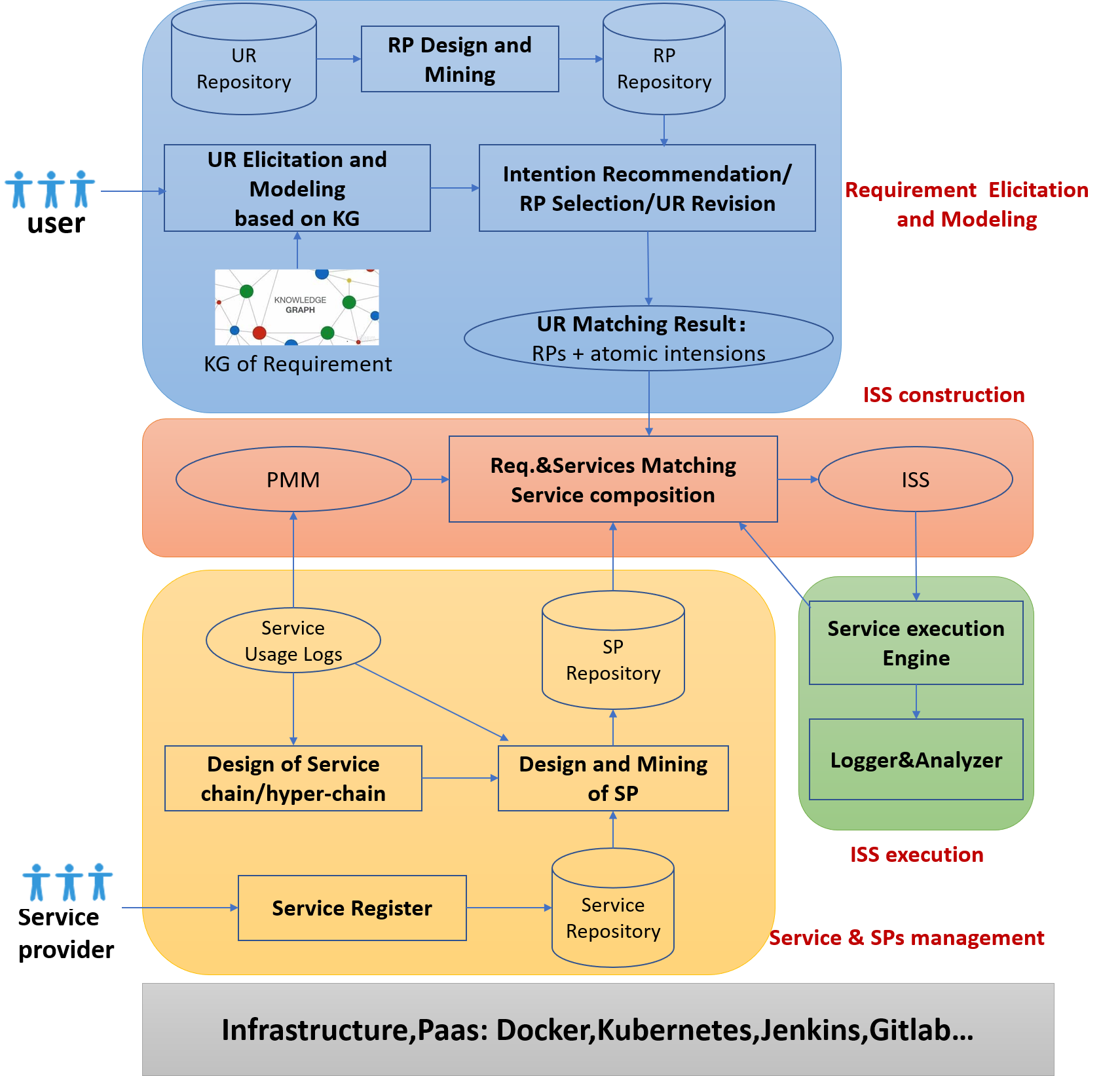}
	\caption{Framework and platform for ISS design of IoS}
	\label{fig:the big picture}
\end{figure}

In our previous works \cite{xu2018new}, a new paradigm of software service engineering for the rapid development of ISSs in the Big Service ecosystem, whose name is RE2SEP (Requirement-Engineering Two-Phase of Service Engineering Paradigm) was proposed. One of the main ideas of RE2SEP is taking advantage of the domain priori knowledge of both the user side and the provider side to derive service patterns and requirement patterns. And then, the probabilistic matching matrices between the bilateral patterns are established to facilitate the service solution construction. Based on these previous works, we propose an ISS design framework and supporting platform of IoS in this paper, the overall architecture of which is shown in Fig. \ref{fig:the big picture}.

%%2020-02-14 关于图中和正文中的术语，建议做以下校正：
%User Requirement, UR, URs
%Requirement Pattern, RP, RPs
%Service Pattern, SP, SPs
%User Intention Tree (I-Tree)
%Probabilistic matching matrix between RPs and SPs (PMM)
%%Integrated Service Solution (ISS)%%最终造出的东西
%Requirement Elicitation and Modeling %%需求捕获和建模，不用presenting
% XHC注:此部分修改完成

Main elements of the framework and platform, and contributions of this paper, are summarized as follows:

\textbf{(1) Service Pattern Management}:
%% The facility for gathering, virtualizing, and organizing real-world services from one or multi-domains is the key foundation of service development and running. Services typically are diverse, distributed, and heterogeneous, so we design an OWL-S based service specification to describe services in a unified format. 
The priori domain usage knowledge of services is used for the construction of Service Patterns (SPs), which refer to typical complete or partial ISSs in a certain domain. The services repository and SP repository are well prepared for the future expected possible ISSs. 

\textbf{(2) Requirement Elicitation and Modeling}: An intention tree model (I-Tree) is proposed, which supports end-users to present requirements in an understandable and convenient way. And then, a knowledge graph of requirements(KGR) is employed to assist users to complete requirement presentation automatically and quickly. It can be found that even various massive customers have relatively limited types of User Requirements (URs), which are the composition of Requirement Patterns (RPs) formed through previous usage experience in certain domains. RPs are used for recommendation and revision of the I-Tree. Finally, optimal RPs are selected to cover the I-Tree as much as possible.  

\textbf{(3) Integrated Service Solution Construction}: A multidimensional probability matching matrix (PMM) is maintained which stores the matching information between SPs and RPs in different contexts. When a specific actual UR is coming, based on the optimal RPs selected in the step of requirement processing, by taking advantage of the PMM, a set of SPs are filtered out and composed together to form the final ISS efficiently and quickly.  

\textbf{(4) IoS Platform}: The supporting platform provides supports to ISS design, development, and running in IoS. It adopts a decentralized and distributed architecture. It supports the organization of multi-layer services and SPs, and can support multi-layer and cross-domain service aggregation.  

The remainder of this paper is organized as follows. Section \ref{Requirement Process} introduces requirement elicitation and modeling based on RP. Section \ref{sec:service} defines the SP. In Section \ref{sec:service solution}, ISSs are constructed based on bilateral patterns and PMM. Section \ref{sec:platform} presents the design of the IoS platform. Section \ref{sec:related work} introduces related work. Section \ref{sec:conclusion} concludes the paper.

\section{Pattern based Requirement Elicitation and Modeling}
\label{Requirement Process}

\subsection{User Intention Tree}
\label{sec:Req Statement}

\begin{figure}[ht]
	\centering
	\includegraphics[width=3in]{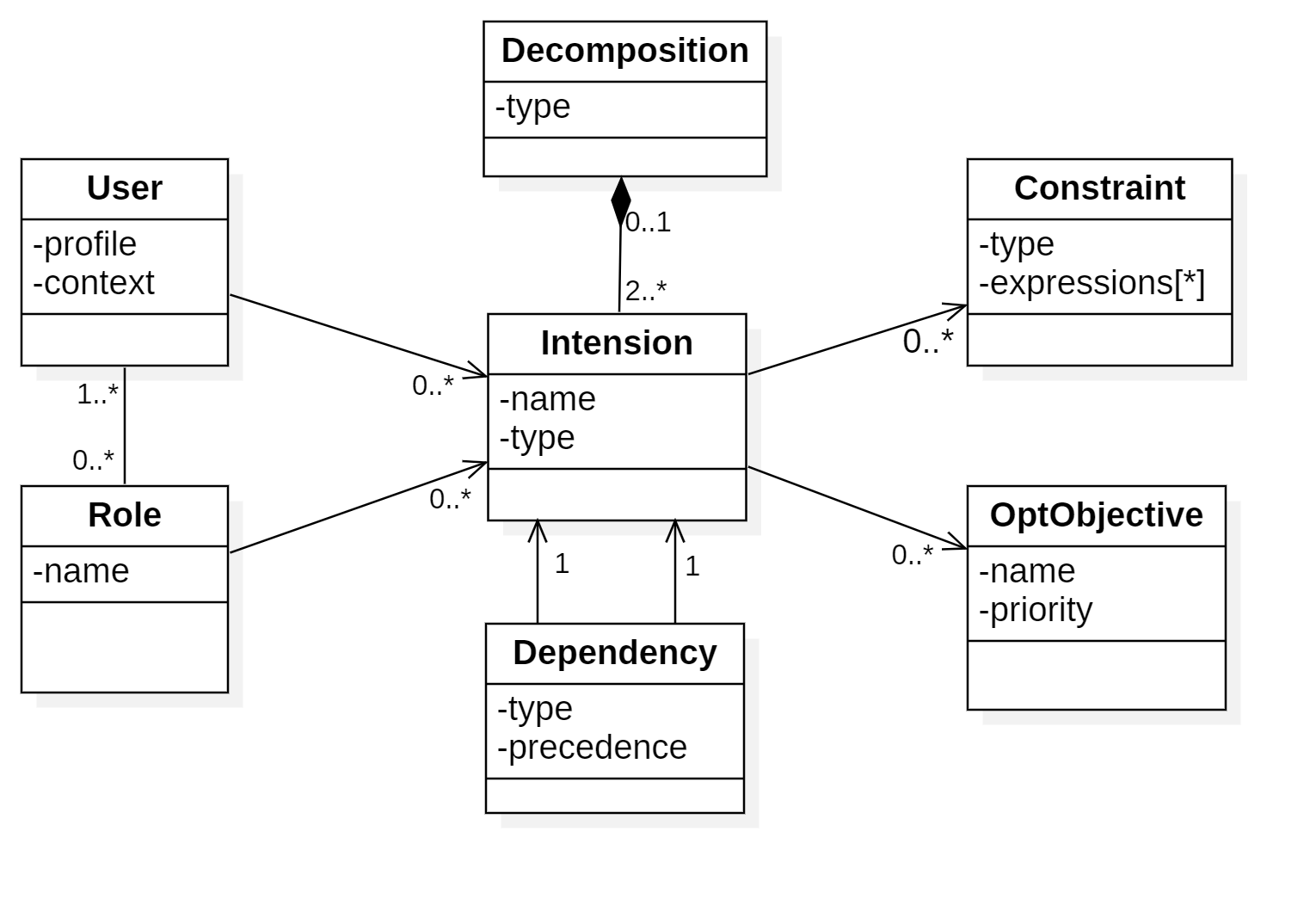}
	\caption{Meta-model of intention tree (I-Tree)}
	\label{fig:I-tree model}
\end{figure}

For end-users in a service system, expressing their requirements in natural language is the most natural and convenient way. However, this way usually leads to ambiguity, which is not conducive to the elicitation and analysis of requirements. In order to elicit and model requirements in an understandable manner and also in a professional well-defined pattern for developers, we propose an intention tree (I-Tree) model based on goal-oriented techniques that are commonly used in requirement engineering and have been advocated to express stakeholder objectives\cite{horkoff2016interactive}. I-Tree model specifies the decomposition and constraint relationships among intentions. Fig. \ref{fig:I-tree model} shows elements and their relationships in the I-Tree meta-model. 

\begin{comment}  %%
The intention tree model is defined as: 
\[ IntenTree= <G,E> \]

%Where $ G $ is the set of goals, $ E=\{(goal_{i},goal_{j})| goal_{i} $ is the parent node of $goal_{j}\} $, $goal= <InDes,Conset>$

$ G $ is the set of goals. $goal= <InDes,Conset>$, where $ InDes $ is the description of the user's intention that express specific functional requirements, represented in natural language. $Conset$ is the set of constraints. $ E=\{(goal_{i},goal_{j})| goal_{i} $ is the parent node of $goal_{j}\} $ which represents the containment relationship between goals. The structure of intention tree is shown in Fig. \ref{fig:structure-intension-tree}.
\end{comment}

As shown in Fig. \ref{fig:I-tree model}, the I-Tree meta-model consists of the following elements: (1) User. (2) Role, which can be played by one or more users, and each user can play one or more roles.. (3) Intention, which describes specific functional requirements of users. (4) Decomposition, which represents a way to further decompose the upper intention into several lower intentions. There are two decomposition relationships between upper and lower intentions: “AND” and “OR”. The “AND” relationship means that the upper intention can be achieved only if all the lower intentions associated with it can be achieved. The “OR” relationship means that the upper intention can be achieved if any one of the lower intentions associated with it can be achieved. (5) Dependency. There are some dependencies between intentions, which means that the realization of one intention depends on another one. (6) Constraint, which denotes the description of the non-functional attributes of the intention. Constraint is defined in key-value pairs. i.e., $Constraint=<Cons_{name},Cons_{type},Cons_{value}>$. Constraints have different types, including enumeration, Boolean, and interval. It is necessary to distinguish different types of constraints because different types of constraints can be dealt with in different ways. (7) OptObjective, which refers to the user's preference for optimization objectives of the intention in ISS construction. The corresponding optimization strategy will be adopted to achieve the intention.

%%2020-02-14 节点应该叫intention。constraints要附着在每个intention节点上，而不是只标注一个。

\begin{figure*}[htbp]
	\centering
	\includegraphics[width=6in,scale=1]{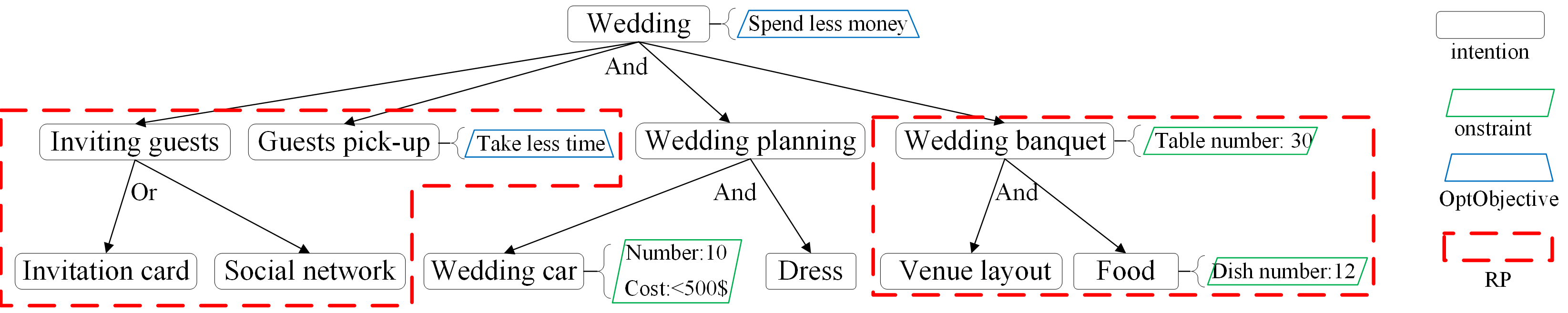}
	\caption{An example of I-Tree and RP}
	\label{fig:example-intension-tree}
\end{figure*}

A practical example of I-Tree model for a wedding service requirement is shown in Fig. \ref{fig:example-intension-tree}. The user plans to hold a wedding. The root intention includes the sub-intentions of the wedding banquet, wedding planning, guest pick-up, and inviting guests. The decomposition type between root intention and its sub-intentions is “And”. The overall optimization objective is the cost as low as possible. In the banquet intention, the user specifies the constraint of the number of tables and decomposes fine-grained intentions: venue layout and food. Similarly, the figure shows users' intention decomposition, constraints, and optimization objectives for the wedding planning, inviting guests, and guest pick-up.

%\begin{figure}[ht]
%%2020-02-14 这个例子感觉太简单了，看不出跨界，看不出对IoS的依赖性，建议复杂一点，层次多些，约束全面些。否则就失去了给例子的意义。

\subsection{Requirement Pattern}

It can be found that there is a set of requirements that always appear together in the user's intention trees. These requirements represent a common routine or relatively stable service process or sub-solution to be reused in satisfying different customer requirements in certain domains. We define these frequently occurring fragments of service requirement as RP, which is a modularized piece of the description of URs. RPs are expected to be reused in satisfying different URs. They are useful for presenting requirements in a quick and efficient way as well as for constructing ISS.

$RP$ is defined as:
\[ RP=<info,\{I-Tree\}> \]
where $ info $ represents fundamental information of the RP, such as frequency of use, domain information, and description information. $ \{I-Tree\} $ is the set of intentions along with their sub-intentions and constraints. An intention in an I-Tree is not only a node but also is a sub-tree. So an RP is a forest composed of I-Trees. In the example of the wedding UR shown in Fig. \ref{fig:example-intension-tree}, there are two RPs, and one is composed of the intentions of inviting and pick-up guest, the other is the wedding banquet intention. RPs have various scales, and some complex ones are composed of I-trees, some may only have one single intention, so that stakeholders can flexibly model requirements via RPs.

To derive RPs from requirement repository, the first step is to mine the frequent substructures from I-Trees whose occur times reach a specific threshold. This can be abstracted as a frequent subgraph mining problem. We design an improved gSpan algorithm to deal with this issue. In the second step, the results, the frequent substructures, are grouped according to their functional requirements. Finally, in each group, intentions are clustered into RPs according to constraints. The constraints of frequent substructures may be different within each group. The intentions are the same if the similarity between the constraints that exceeds the threshold. By comparison, frequent substructures can be divided into several classes according to the constraints. RP can be clustered from these similar frequent substructures by keeping common constraints and removing personalized ones. We will present the detailed RPs deriving method in another article. 

\subsection{Overall Process of Requirements}

\begin{figure}[ht]
	\centering
	\includegraphics[width=3.5in, height = 2.1in]{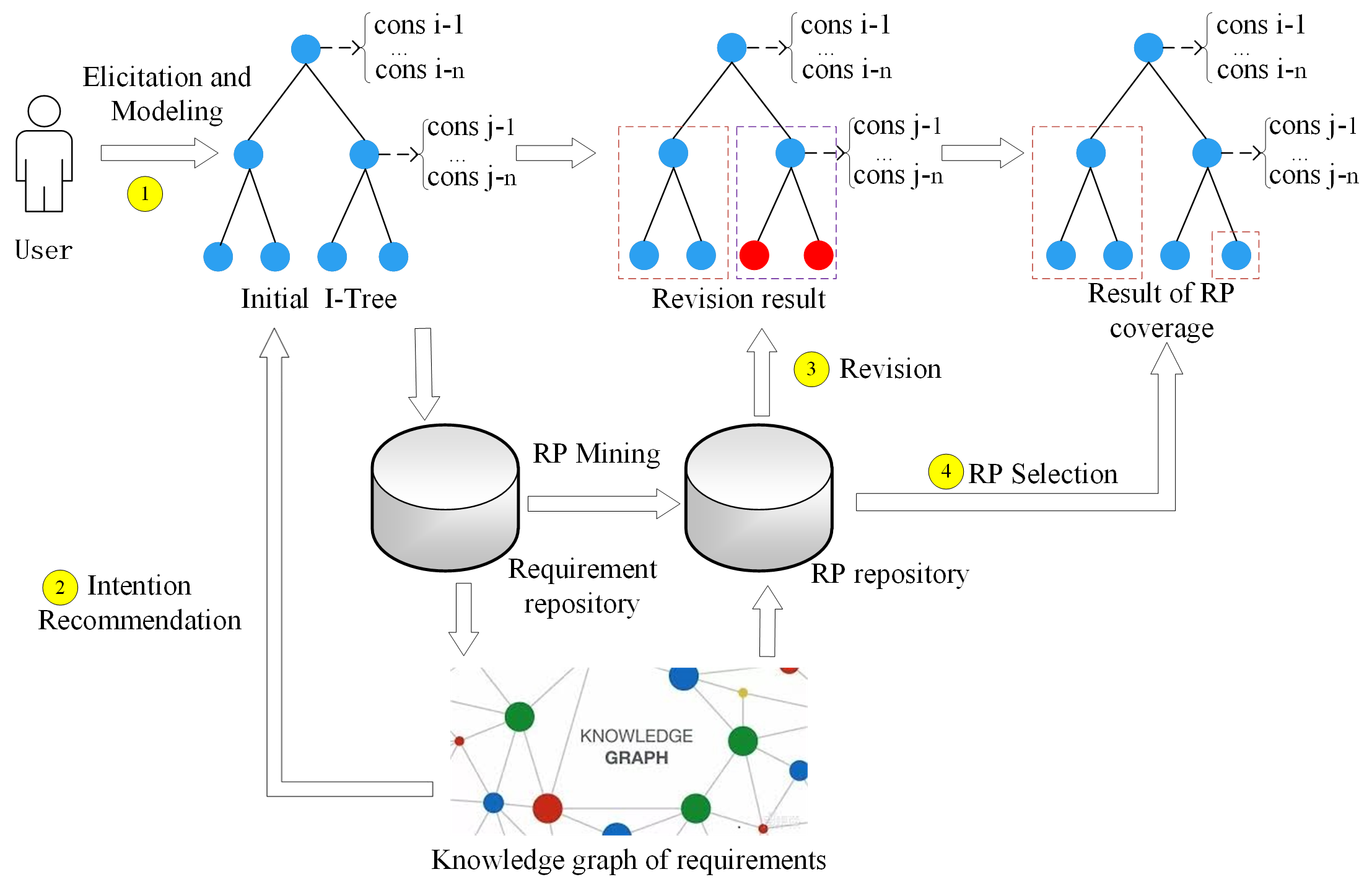}
	\caption{Requirement processing}
	\label{fig:requirement-processing}
\end{figure}

%%2020-02-14 不建议上来就给出图并解释。得需要先讲以下道理，应对introduction里你提到的关于需求层面的挑战。为什么要做KG-based recommendation，为什么要选择RP覆盖，为什么要revision，从而凸显咱们方法的创新性。
To address the challenge of the requirement process described in Section I, we first propose the I-Tree model to elicit and model URs clearly and conveniently. Then a
KGR is generated and exploited to recommend related intentions to users for assisting them to complete the I-Tree automatically. Finally, RPs are selected to cover I-Trees. In the solution construction phase, corresponding SPs will be selected based on PPM to construct a solution meeting the URs quickly. The overall requirement process is shown in Fig. \ref{fig:requirement-processing}.
%%2020-02-14 
% Obtainment --> Elicitation and Modeling
% Completion --> Intention Recommendation 本意说的是：利用外部知识图谱，根据用户已建立的部分需求，来推荐其他意图进去。用Completion不适合。 
% Matching  ---> RP Selection 我们平常叫匹配，但其实很不准确。本意应该是选择RP来覆盖需求。
% Rewriting ---> 直接用rewriting会让人费解，不如直接点，叫Revision，根据RP，建议用户对需求作何种调整。
% Discovering of RP --> RP Mining
% Pository --> Repository
% match result--> Result of RP coverage
% suggest result--> Revised UR

%%2020-02-14 这个图里的所有文字，第一个字母应该大写。另外该图高和宽比例有些失调。

\textbf{Step1: Requirement Elicitation and Modeling}
At the beginning of the requirement elicitation, users model I-Trees with their intentions and corresponding constraints, and initial I-Tree models are constructed. It’s worth noting that there are two special situations of users' operation in the modeling process. One is that users may be reluctant to spend too much time filling in the intentions they need but don't care about. So they may not fill I-Trees in completely. The other is that users don't need some sub-intentions, so they also don't fill the sub-intentions into the I-Trees. It is hard to distinguish between these two situations. The rule we adopted is to consider only the intentions filled in by users, i.e., considering the second situation. In the step of intention recommendation described below, we will recommend intentions to users to help them complete the I-Trees automatically to deal with the first situation.  

\textbf{Step2: Intention Recommendation}
To assist users to model I-Trees, we design an intention recommendation algorithm based on the KGR. The intention recommendation algorithm can recommend frequently-used related intentions to users or help users fill in incomplete requirements by generating candidate intentions, just like the search candidates provided by a search engine.

The KGR is generated from URs stored in the requirement repository. When users enter requirements, the intention recommendation algorithm can deduce several possible intentions in the KGR based on users’ input and context information in I-Trees. Then the recommendation results are sorted according to specific rules, such as the similarity between requirements and user input, the use frequency of requirements. Finally, the sorted intention list become candidates provided to users for completing I-Trees.

\textbf{Step3: Revision}
In the previous two steps, users fill in I-Trees based on their wishes and knowledge. They may not be familiar with the high-quality RPs and services that IoS platform can provide. So the platform may not be able to provide the best ISS based on the current I-Trees. For such I-Trees, We design a revision approach to generate new similar ones by fine-tuning the users’ intentions according to the RPs with high popularity in the RP repository or relaxing the constraints. These new I-Trees will feedback to users as references. Users can modify the original I-Trees referring to them or accept the revision ones.

\textbf{Step4: RP Selection}
After getting the final I-Tree, the RP selection algorithm is exploited to find the RPs, whose intentions can cover intentions of the I-Trees partially or completely. Here “cover” means that the intentions in RP are the same as the intentions of the I-Tree, and the constraints of the former are looser than the latter. The selection algorithm is run iteratively until any RPs can’t cover the rest uncovered intentions of the I-Tree. There may be more than one selection schemes and the best one should be selected as the final selection result for generating an ISS. The rule is that the selection scheme with the highest coverage of the I-Tree and the smallest number of RPs is the best.

\section{Unified Service Specification and Service Pattern}
\label{sec:service}

\subsection{Service Pattern}
Service pattern(SP) refers to a complete or partial ISS that is frequently used in a specific domain and has a business process, service activities, and service resources.

There are three essential elements of SP. The first is a description of the service process. The second is a set of service instances, which is related to the service business process and can be used to instantiate the SP. The last is the verifying degree of the SP, that is, the evaluation index of SP. It includes the reusability of SPs and the promotion effect of optimization efficiency like increasing the matching speed or enhancing the optimization objectives., etc.

SP reflects the apriority of service in business and optimization. It is a service complex composed of behavior based on business association, and it is a combination in essence. SPs are often extracted by analyzing a large number of business experiences, and they can also be defined by domain experts.

We can formally define an $SP$ as:
\[ SP=<info,fr,process,QoS,cons,instances> \]
where $ info $ is the basic information of the SP, $ fr $ is the functional requirement of the SP, $ process $ is used to describe the service process information of the SP, we use BPMN (Business Process Model and Notation) 2.0 XML format to describe and store the service process, $ QoS $ is the quality information of the SP, $ cons $ is the execution constraint information of the SP, and $ instances $ is the instance set of the SP.

For example, taking Uber and taking a taxi are two services that have the same function: urban traffic. So they can be looked on as the same kind of service. Alike, traveling by train and traveling by air are also the same kind of service. Historical service usage data show that taking Uber and traveling by train are often composited together at the same time in an ISS to meet URs, so taking a taxi and traveling by air do. So $<$urban traffic, inter-city traffic$>$ can be abstracted as an SP, and $<$taking Uber, traveling by train$>$ and $<$taking a taxi,  traveling by air$>$ are two SP instances.

Compared with using atomic services, using SPs to construct ISSs has multiple advantages, such as to achieve fast matching of URs, to build ISSs quickly, and to improve the efficiency of service composition optimization to get better ISS. Due to the large granularity of SPs, an SP can meet the requirements of users in multiple service activities and can get ISSs faster and get better user satisfaction.

We use k-means clustering algorithm, and frequent sub-graph mining approaches gSpan to derive SPs. Firstly, we group similar services into multi-dimensional groups according to the multi-dimensional similarity measurement of function, input and output, user group, and provider, and calculate the conditional association probability of each dimension group. Then, aiming at the characteristics of services, we use frequent sub-graph mining algorithm to recognize the high-frequency service segments. Finally, the SP is abstracted from the SP instances by semi-supervised clustering to ensure that the instances in each class cluster have the same process workflow, and the service class in the instances are the same.

\begin{comment}

\begin{figure}[ht]
	\centering
	\includegraphics[width=3.5in]{figures/example_of_mining_SP.png}
	\caption{Example of mining service pattern.}
	\label{fig:mining SP}
\end{figure}
\end{comment}

The recognition of SP focuses on two key indicators: granularity (the number of service activities included in SP) and usage frequency (support). Granularity determines the reusability of the SP. Small-grained SPs have higher reusability, but lower reuse value; large-granularity SPs have lower reusability, but high reuse value. Therefore, the granularity of SP needs to be balanced between reusability and reuse value. The usage frequency determines the possibility of using SPs, and the choice of frequency should also be balanced between the number of SPs and the frequency. Finally, we extract different granularity and frequency SPs to complete the fast matching of supply with demand.

\section{ISS Construction Based on Bilateral Patterns}
\label{sec:service solution}

\subsection{Problem Description and Model}
The essence of ISS construction in IoS is a supply-demand problem, i.e., matching the requirements from users(demand side) and the supplies from service providers(supply side) fast and accurately, which is a typical complex optimization problem. Since the minimization problem and the maximization problem can be transformed equally, its general model can be described as follows:
\begin{gather}
    \min F(\bar{x})=(f_{1}(\bar{x}),\ldots,f_{p}(\bar{x}))^{T}~~
\end{gather}
\centerline {s.t.}    
\begin{gather}
    g_{i}(\bar{x})\leq 0, ~~i=1,\ldots,k\\
    h_{j}(\bar{x})=0, ~~j=k+1,\ldots,m\\
    \bar{x}\in R^{n}
\end{gather}

where 
\begin{itemize}
  \item $F(\bar{x})=(f_{1}(\bar{x}),\ldots,f_{p}(\bar{x}))^{T}$ is objective function vector, i.e., $p$ objectives to be optimized. When $p=1$, It's a single-objective optimization problem. The optimization objectives of service systems usually include the shortest service completion time, the lowest user expenditure cost, the largest profit of the service provider, the largest resource utilization, the highest user satisfaction, and so on.
  
  \item $g_{i}(\bar{x})\leq 0, (i=1,\ldots,k)$ and $h_{j}(\bar{x})=0, (j=k+1,\ldots,m)$ are constraints. In general, the constraints come from the demand side and the service supply side. Demand deadlines, precedence constraints between service tasks, response time, demand budget, etc. are demand constraints. Service resource available quantity, supply modes(Reserved, On-demand, Spot), etc. are supply constraints.
  
  \item $ \bar{x}\in R^{n}$ are decision variables.
\end{itemize}

\subsection{ ISS Construction Algorithms} 
ISS construction is a kind of very complex optimization problem with characteristics of large scale, multi objectives, and complex constraints. These problems are basically NP-hard problems. Under the existing conditions, it is difficult to design new algorithms and increase computing resources to solve the problem effectively. Our idea is modularizing fine-grained requirements and services into coarse-grained RPs and SPs based on priori domain knowledge and then using intelligent algorithms to reduce the scale of the original problem. The RPs and SPs reflect the fixed collocation of requirements and services. Therefore, a complex requirement can be decomposed into different sub-modules(RPs), and the whole problem is decomposed into some sub-problems. The sub-problems can be further decomposed until all requirements can be achieved by SPs or atomic services.

The SPs are composed of the services that frequently serve the requirements together, but the same RPs in different contexts and with different optimization objectives can be matched by different SPs. How to choose the appropriate SPs corresponding to the RPs is the core of the ISS construction, which needs to reflect the effectiveness and efficiency of the algorithm. Via decomposition into RPs, a large-scale problem with complex UR can be decomposed into some small-scale sub-problems to improve efficiency; meanwhile, SPs are the optimization results obtained through machine learning approaches from priori domain knowledge, and reasonable matching can ensure the effectiveness of the algorithm. 

\begin{figure}[htbp]
	\centering
	\includegraphics[width=3.5in,height=1.5in]{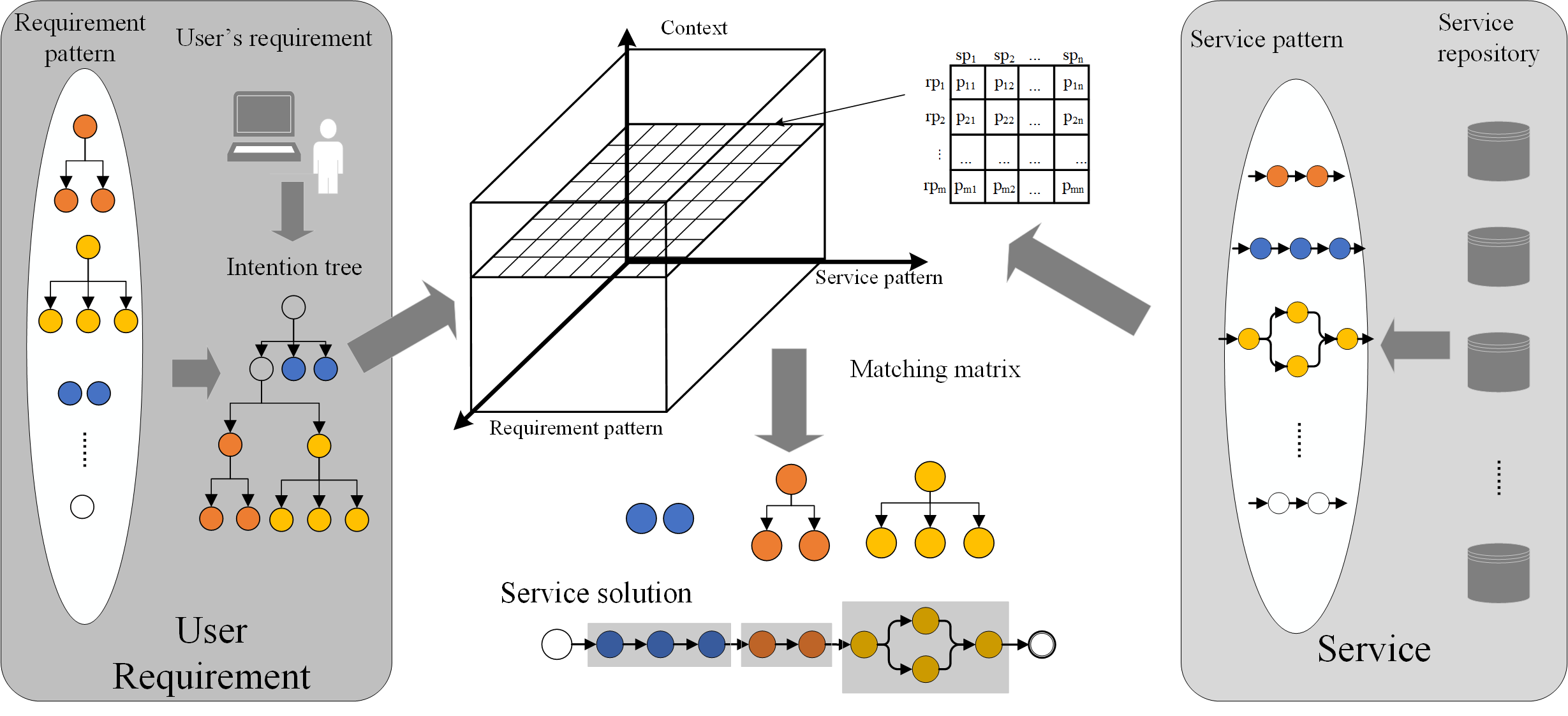}
	\caption{The matching matrix (PMM) and matching process}
	\label{fig:matching matrix and process}
\end{figure}

To further improve matching efficiency and success rate, we construct a probability matching matrix(PMM) to link RPs and SPs. The PMM is a description of the relationship between RPs and SPs, and constructed by summarizing successful experiences and domain knowledge. Because in different service contexts and for different optimization objectives, an RP can be matched by different SPs with different probabilities, the PMM is multi-dimensional, as shown in Fig. \ref{fig:matching matrix and process}. The X-axis is the RPs. The Y-axis is the SPs, and the Z-axis is the context that describes the matching scenarios. Context includes the information of users, environment, and objectives.

%这个图的比例也失调

%Suppose there are $p$ contexts. Let $b_{k,j}$ be the matching probability in context $k$ when $SP_j$ is selected to achieve $RP_{i}$. $0\leq b_{k,j}\leq 1$. For simplicity, let's assume that each context has $m_{0}$ candidate SPs. The corresponding PMM can be expressed as sparse matrix $\mathcal{B}^{i}$.:
%
%\begin{equation*}
%    \mathcal{B}^{i}=\begin{bmatrix}
%    b_{1,1}&b_{1,2}&\ldots&b_{1,m_{0}}\\
%    b_{2,1}&b_{2,2}&\ldots&b_{2,m_{0}}\\
%    \ldots&\ldots&\ldots&\ldots\\
%    b_{p,1}&b_{p,2}&\ldots&b_{p,m_{0}}
%    \end{bmatrix}
%\end{equation*}

The matching probability is calculated by a comprehensive evaluation of using times, matching quality results, matching difficulty, and other factors. PMM will be updated periodically. 

We have designed and implemented a series of algorithms based on PMM for different contexts and optimization strategies. The algorithms are categorized for:
\begin{itemize}
  \item Number of objective: \textit{single}, \textit{multi}
	\item Priority of solution quality and efficiency: \textit{the optimal solution}, \textit{the satisfactory solution}
	\item Service supply modes: \textit{reserved}, \textit{on-demand}, \textit{spot}
	\item Algorithm strategy: \textit{exact}, \textit{rule-based}, \textit{heuristic}, \textit{meta-heuristic}
\end{itemize}

\section{Design and Implementation of IoS Platform}
\label{sec:platform}
IoS platform needs to aggregate various services from various organizations and domains. To this end, a decentralized and distributed architecture is adopted in the Big Service platform, which means each node has the same functions and implementations as peers. %In the future, we hope that each organization could develop its own customized IoS nodes just following the same external interface.

IoS nodes use the P2P protocol for discovery and routing, and use inter-node communication to achieve cross-node requirement service matching and service solution execution. Platform users can post requirements, perform service matching, generate solutions, and execute on any node. Each node not only allows services to be deployed inside, but also allows registration of external services and third-party platforms.

\begin{figure}[ht]
	\centering
	\includegraphics[width=3.25in]{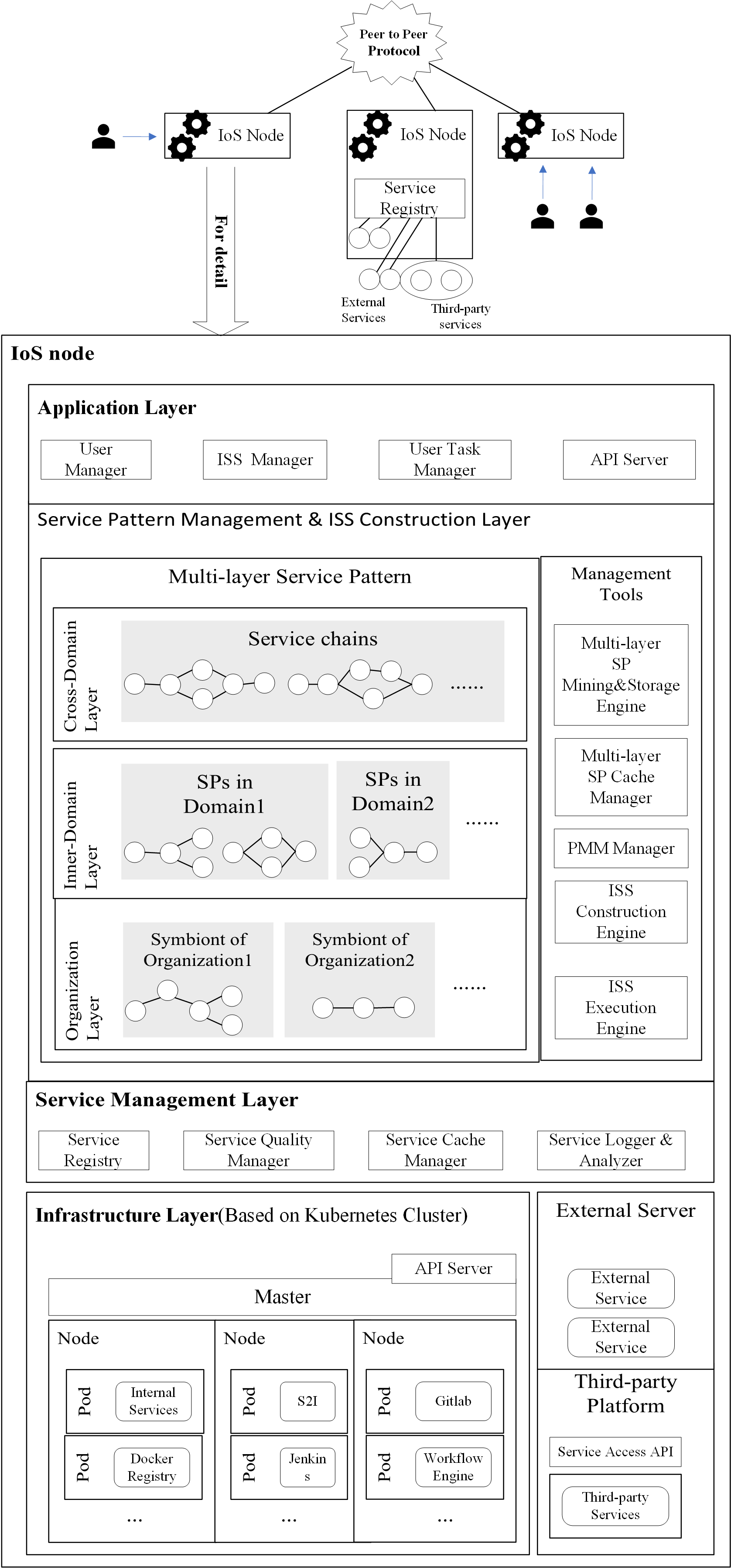}
	\caption{Architecture of the platform}
	\label{fig:architecture}
\end{figure}

The architecture of each IoS node is shown in Fig. \ref{fig:architecture}, which can be divided into four layers.
%%2020-02-13 感觉这个图这么画意义不大，只是把之前各章节介绍的功能模块汇集到了一起。建议（1）可否把技术架构层面表达进去，展现出云、分布式服务、第三方平台、分散的用户等。（2）把各层服务模式的生态呈现出来，从而扩展前面的SP：我们的平台是多层SP。（3）底层的infrastructure可以扩展展示出细节来，表明我们的平台用了什么新的技术。（4）除了这张图，是否应给出几个代表性的UI，表明我们做出来了，而不是仅仅设计出来了。

%%

\subsection{Infrastructure Layer}
Infrastructure Layer provides underlying supports such as container construction, management, and deployment in the cluster. We use Docker as the underlying containers of the platform and use Kubernetes to manage the containers. The Docker Registry, automatic image build tools(S2I and Jenkins), performance monitoring tools, and the services deployed on the IoS platform all run in the Kubernetes pods. Docker Registry is deployed in the cluster to store service images. The S2I and Jenkins provide two different ways to build Docker images from source code.

\subsection{Service Management Layer}
Service Management Layer provides supports for service management such as service registration, service quality management, service information caching, service log collection and processing.

\subsection{Service Pattern Management \& ISS Construction Layer}
This layer consists of the multi-layer service pattern mining and storage tools, and ISS construction and execution engines.

As introduced above, SPs are harvested from priori domain knowledge. SPs are organized into three layers, i.e., the organization layer, the Inner-Domain layer, and the Cross-Domain layer. In the organization layer, we aggregate the cooperation relationships between the services provided by the same service provider. There are two types of entities, i.e., atomic services and service patterns. The relationships between entities in this layer are mainly cooperative relationships, which contain four sub-types as follows: symbiosis, parasite, location complementarity, and temporal complementarity. In the Inner-Domain layer, services and SPs from different organizations in the same domain are aggregated to a service chain that spans multiple organizations. Furthermore, after further grouping according to the service features conversion layer entities, the competitive relationship between the entities can be extracted. In the Cross-Domain layer, SPs that cross different domains are linked together to form a service super-chain, which corresponds to specific coarse-grained requirements.

\subsection{Application Layer}
\begin{figure}[ht]
	\centering
	\includegraphics[width=3.25in]{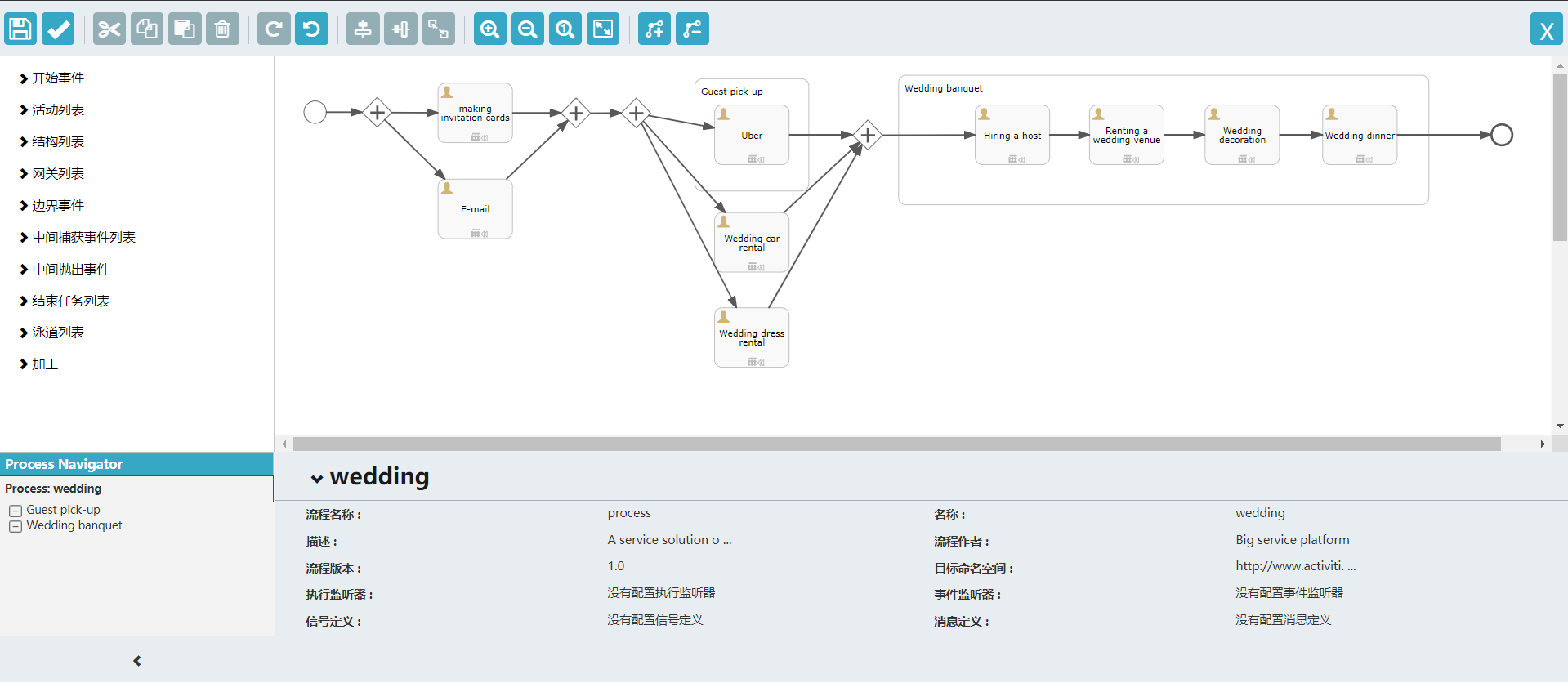}
	\caption{System interface example: ISS construction}
	\label{fig:exapmple1}
\end{figure}

We provide interactive interfaces and management applications to users. One example of ISS construction is shown in Fig. \ref{fig:exapmple1}.

%\begin{figure}[ht]
%	\centering
%	\includegraphics[width=3.25in]{figures/system-interface-2.jpg}
%	\caption{System interface example: ISS construction}
%	\label{fig:exapmple2}
%\end{figure}

\begin{comment}
\begin{itemize}
	\item Infrastructure Layer, which provides underlying supports such as container construction, management, and deployment in the cluster.
	\item Service Management Layer, which provides supports for Service Management such as service registration, service quality management, service information caching, service log collection and processing.
	\item Service Pattern Management & ISS Construction Layer, which provides support for multi-layer service patterns. In this layer, the multi-layer service pattern mining, storage, and ISS construction and execution are implemented.
	\item Application Layer, which provides applications for users.
\end{itemize}
\end{comment}

%%2020-02-14 相关工作这一节里，感觉缺了一个对比分析。虽然我们没法作实验，但还是要有定性的对比分析。给出一张表格，从多个侧面比较我们的工作和其他研究/实践工作的差异，凸显我们的创新。
%% 与传统的Reuse-based Software Engineering (RBSE) / Software Product Lines (SPL)的对比，RBSE/SPL也是两阶段：需求工程、领域工程，然后匹配。
%% 与传统服务选择、服务组合、服务推荐的对比：从0还是非0开始，小粒度方案还是大粒度方案、跨领域还是单一领域
%% 与第三方平台例如携程淘宝等的对比：我们的平台是跨领域的，应对跨界的需求；动态构建方案+长久存在的稳定方案一起work、...

%%2020-02-14 通过对比所要凸显的创新
%%  从0开始还是从一个基础开始？凸显我们的领域先验的价值
%%  需求的表达：意图树的灵活性、自由性、使用简便性、描述能力强、与知识图谱连接，可智能化推荐/修订
%%  匹配：凸显效率
%%  服务库和服务模式库：凸显跨领域的聚合、多层生态性，逐层聚合
%%  ...其他的我再考虑

\section{Related Work}% and Qualitative Comparison}
\label{sec:related work}
\subsection{Pattern-based Service Construction}
Service patterns are gradually extended from traditional workflow patterns to service matching, and in most related studies, patterns are regarded as structured workflow or business process. \cite{Xu2017RE2SEP} defines the SP and proposes that RPs are matched with SPs with service context as mediation. 
%In the early days of service computing, SP is regarded as an abstract template, which can be instantiated in different forms\cite{Melloul2004Reusable}. 
It is found that historical ISSs contains much priori service usage knowledge, which have a great impact on the construction of ISS, so mining patterns from historical ISSs with the data-mining techniques becomes the main approach\cite{Wang2011Mining}. \cite{upadhyaya2013approach} extracts the frequent process fragments in service composition through data-mining techniques, and use the fragments in service composition that followed. %Huang\cite{Huang2010Business} gives a method of composite service selection using the mined service process fragments.
\cite{Liu2017Probability} uses the FP-growth mining method to mine valuable SPs from historical ISSs. These schemes try to automatically acquire the business patterns that are reused in history, identify the similar parts in structure and function, and combine them as high-level abstract patterns.

\subsection{Goal-oriented Requirement acquisition}
Requirement acquisition is a classic topic. Many scholars use ontology-based analysis to explore traditional requirements acquisition methods. \cite{souag2015security} used ontology to obtain requirements for security software. \cite{chen2016capturing} proposed an ontology-based approach to analyze the interaction between software and its interactive environment to capture software requirements. With the accumulation of data, many scholars use the method of data analysis to construct user portraits to predict user requirements. \cite{li2014weakly} adopts the method of weakly supervised learning to extract attribute labels from Twitter social data to construct users' portraits. Besides, much research has been done by extracting entities and generating associations between entities and users' interests to realize the focus mining of domain users\cite{qiang2014service}.% extracted the theme model related to the service description of the Internet of things for the selection of user services by mining the library content such as Wikipedia. \cite{Gao2016SeCo} mined service event topics for service recommendation through subject analysis of API description text and user feedback from ProgrammableWeb.com, an open-source Web services community.

\subsection{Related Frameworks and Comparison}

Because the core idea of this paper is to mine bilateral patterns, RPs and SPs, from priori domain knowledge for reuse in a large-grained way, several related frameworks based on reuse techniques are selected for comparison. Software Product Line (SPL) engineering refers to the engineering and management techniques to create, evolve, and sustain a software product line which is a portfolio of similar software-based systems and products produced from a shared set of software assets using a common means of production\cite{bastarrica2019software}. Reuse-based Software Engineering (RBSE) promotes productivity by avoiding redevelopment and improves quality by integrating components that have established reliability\cite{selby2005enabling}\cite{perucci2017multipurpose}. Service-Oriented Modeling and Architecture(SOMA) implements a component-based software engineering approach to service-oriented analysis and design, including service identification, service definition, and service implementation\cite{arsanjani2008soma}. Service Model Driven Architecture(SMDA) helps service engineers describe the design and implementation of service systems that meet user service requirements through a model-driven approach\cite{xu2007smda}. Service-Oriented Development In a Unified framework (SODIUM) presents a model for developing service applications, focusing on how to define new services using coarse-grained combinations of services that can be reused\cite{topouzidou2007service}. 

We compare the related frameworks with our proposed framework in the following aspects:1)  Service Constructon Stages, how the services are constructed. 2) Requirement Utilization, in requirement elicitation and modeling, what kinds of requirements are employed, i.e., atomic or/and composite requirements(RP). 3) Service Utilization, in service selection and composition, what kinds of services are employed, i.e., atomic or/and composite services(SP). 4) Domain Service Supported, whether supporting the construction for single-domain servcies or/and for cross-domain services. The comparison result is shown in Table \ref{tbl:Comparison of different service developing paradigms}. 

The service construction process of RBSE, SPl and our proposed framework can be divided into three main stages. The first stage is called Requirement Engineering(RE), which is a top-down process and elicitating and modeling URs. The Second stage is the Domain Engineering(DE), which is a bottom-up process with the task of finding and developing reusable services. The last stage is to construct service solutions based on results of RE and DE. Other framworks are top-down one-way processes. With the With the bidirectional strategy of RE-DE, domain knowledge and service assets can be better utilized on both the demand side and the service provsion side. Moreover, only our framework provides the supports of mining and utilizaiton of bilateral patterns, which can make full use of priori domain knowledge on both the demand side and the service side to improve the construction efficiency and user satisfaction of service solutions. It is worth pointing out that RPs and SPs can be used in other frameworks, we assign the values of those fields to star to indecate this in Table \ref{tbl:Comparison of different service developing paradigms}. Last, as described in Section V-C, SPs are harvested from priori domain knowledge within organizations, within domains, and across domains, so our proposed framework supports the construction of cross-domain services. To conclude, the proposed framework makes better use of domain priori knowledge derived from the commonness and similarities among massive historical URs and among historical ISSs, which leads to it can provide supports for the quick and accurate construction of ISSs.

%% Please add the following required packages to your document preamble:
%% \usepackage{multirow}
%\begin{table*}[ht]
%	\centering
%	\begin{tabular}{|c|c|c|c|c|c|c|}
%		\hline
%		\multicolumn{1}{|c|}{\multirow{2}{*}{Approaches}} & \multirow{2}{*}{Service Construction Stages} & \multicolumn{2}{c|}{Criteria} \\ \cline{3-7} 
%		\multicolumn{1}{|c|}{}   &  & Req. model & Domain knowledge & Stages & Domains  & Supporting platform  \\ \hline
%		RBSE & SSE paradiam   &      &      &     &     &     \\ \hline
%		SPL & SSE paradiam  &      &      &     &     &     \\ \hline
%		A1 & ISS construction algorithm  &      &      &     &     &     \\ \hline
%		A2 & ISS construction algorithm  &      &      &     &     &     \\ \hline
%		A3 & platform  &      &      &     &     &     \\ \hline
%		the proposed approach & Multi-perspectives  &      &      &     &     &     \\ \hline
%	\end{tabular}
%\end{table*}

\begin{table*}[ht]
	\caption{Comparison of different service developing paradigms}
	\label{tbl:Comparison of different service developing paradigms}
	\centering
	\begin{tabular}{c|c|c|c|c|c|c|c}
		\hline
		
		\multicolumn{1}{c|}{\multirow{2}{*}{Frameworks}} & \multirow{2}{*}{Service Construction Stages} & \multicolumn{2}{c|}{Requirement Utilization} &\multicolumn{2}{c|}{Service Utilization} &\multicolumn{2}{c}{Domain Service Supported}\\ \cline{3-8} 
		
		\multicolumn{1}{c|}{}   &  & atomic Req. & RP & atomic service & SP & single-domain & cross-domain\\ \hline
	
		RBSE& RE-DE,top-down\&bottom-up &+&*&+&*&+&- \\ \hline
		SPL& RE-DE,top-down\&bottom-up &+&*&+&*&+&- \\ \hline
		SOMA& top-down&+&*&+&*&+&- \\ \hline
		SMDA& multi-stages,top-down&+&-&+&-&+&- \\ \hline
		SODIUM& multi-stages,top-down&+&-&+&-&+&- \\ \hline
		The proposed framework& RE-DE,top-down\&bottom-up&+&+&+&+&+&+ \\ \hline
		%几个部分支持的
		
	\end{tabular}
\end{table*}
% column 1: RBSE、SPL、A1、A2、A3、The

%
%\begin{table*}[ht]
%	\caption{Comparison of different service selection/composition approaches}
%	\label{tbl:Comparison of Comparison of different service selection/composition approaches}
%	\centering
%	\begin{tabular}{c|c|c|c|c|c|c|c}
%		\hline
%	
%		Approaches& From scratch & From  &Fine-grained service&Coarse-grained service&-&+&- \\ \hline
%		
%		The proposed approach & two stages&+&-&+&-&+&- \\ \hline
%		SPL& two stages&+&-&+&-&+&- \\ \hline
%		SPL& top-down&+&-&+&-&+&- \\ \hline
%		SPL& top-down&+&-&+&-&+&- \\ \hline
%		SPL& top-down&+&-&+&-&+&- \\ \hline
%		The proposed apporach& two stages&+&-&+&-&+&- \\ \hline
%		%几个部分支持的
%		
%		
%		
%	\end{tabular}
%\end{table*}

\section{Conclusion}
\label{sec:conclusion}
To quickly and efficiently construct ISSs in IoS, we present a service designing framework and supporting platform of IoS. The framework takes advantage of priori usage knowledge of both the customer side and the provider side to derive SPs and RPs. And then, the probabilistic matching matrices between bilateral patterns are established. When a new requirement is coming, designers or customers can present it easily and explicitly with the proposed intention tree model and RPs. Subsequently, based on the bilateral patterns and their matching matrices, ISSs can be gotten more efficiently than traditional approaches. Finally, the introduced platform can provide the necessary management and running supports for the framework.

Our ongoing and future work includes: (1) to design more supply-demand matching optimization algorithms to support the service construction with various optimization objectives. (2) More application case studies will be investigated and performed to verify the effectiveness of the framework and the performance of the platform. 

\section*{Acknowledgment}
Research works in this paper are supported by the National Key R\&D Program of China (No. 2018YFB1402500) and the National Natural Science Foundation (NSF) of China (Nos. 61832004, 61772159, 61832014).



%\bibliographystyle{IEEEtran}

%\bibliography{hanchuan}
%\begin{thebibliography}{1}
%\end{thebibliography}
%

\bibliographystyle{IEEEtran}
\bibliography{Hanchuan.bib}

\end{document}